\documentclass[12pt]{JHEP3}
\usepackage{amssymb,amsmath,amsfonts}
\usepackage{cite}

\title{Graviton 1-loop partition function for 3-dimensional massive gravity}

\author{Matthias R.\ Gaberdiel\\
           Institut f\"ur Theoretische Physik, ETH Z\"urich\\
            8093 Z\"urich, Switzerland\\
           Email: \email{mrg@itp.phys.ethz.ch}}

\author{Daniel Grumiller\\
           Institute for Theoretical Physics, 
           Vienna University of Technology,\\
           Wiedner Hauptstr. 8--10/136,
           A-1040 Vienna, Austria\\
           Email: \email{grumil@hep.itp.tuwien.ac.at}}

 \author{Dmitri Vassilevich\\
          CMCC, Universidade Federal do ABC,\\
          Rua Santa Ad\'elia, 166,
          Santo Andr\'e, SP BRAZIL \\
          and\\
          Department of Theoretical Physics, St Petersburg State University,\\
          St Petersburg, Russia\\
          Email: \email{dvassil@gmail.com}}

\abstract{
The graviton 1-loop partition function in Euclidean topologically massive gravity (TMG)
is calculated using heat kernel techniques. The partition function does not factorize holomorphically,
and at the chiral point it has the structure expected from a logarithmic conformal field theory. 
This gives strong evidence for the proposal that the dual conformal field theory to
TMG at the chiral point is indeed logarithmic. 
We also generalize our results to new massive gravity. 
}


\preprint{TUW--10--11}

\begin{document}

\section{Introduction}

There are many reasons to believe that quantum gravity will have its simplest realization in 
negatively curved AdS spaces. 
For instance, gravity in positively curved de Sitter space is most likely metastable at best. 
Furthermore AdS gravity is often dual to a conformal field theory (CFT) which defines a 
well-defined quantum system. 
In 2007 Witten exploited the intriguing idea of finding a CFT dual to 3-dimensional 
Einstein gravity with negative cosmological constant, and conjectured this dual CFT to be 
extremal \cite{Witten:2007kt}. Subsequently, Maloney and Witten calculated the 
gravtion 1-loop partition function of this theory \cite{Maloney:2007ud}, but their result did not factorize 
into left- and right-moving contributions, thereby violating one of the assumptions of the original proposal
\cite{Witten:2007kt}. It also did not seem to give rise to a sensible CFT partition function.
The original proposal of Witten was furthermore in conflict with a 
conformal field theory argument that suggested that extremal CFTs cannot exist for large central 
charges \cite{Gaberdiel:2007ve}.

Soon after Witten's proposal Li, Song and Strominger  \cite{Li:2008dq} considered a slightly modified 
version  of Witten's setup by replacing Einstein gravity with `chiral gravity' \cite{Li:2008dq}, i.e.\
topologically massive gravity 
(TMG) \cite{Deser:1982vy,
Deser:1982sv} for a specific value of the coupling constant. The distinguishing feature of this theory
is that  one of the two central charges vanishes, while the other is still non-zero. 
This suggests that the partition function  might factorize into a trivial left- and a non-trivial 
right-moving contribution \cite{Maloney:2009ck}.
TMG is a 3-dimensional theory of gravity with many intriguing features. 
For example, TMG exhibits massive gravity waves, black hole solutions \cite{Banados:1992wn}, 
solutions that asymptote to AdS as well as solutions with different asymptotic behavior, like 
squashed/stretched AdS or spacetimes with asymptotic Schr\"odinger scaling. 
Many interesting results have been obtained at the classical level, see 
e.g.~\cite{Chow:2009km} 
and references therein for a summary of exact solutions, and \cite{Ertl:2010dh} for all stationary 
axi-symmetric solutions. 

Again, soon after the proposal of Li, Song and Strominger  \cite{Li:2008dq},  it was conjectured that the
dual CFT for TMG at the chiral (or critical) point is in fact not chiral, but logarithmic 
\cite{Grumiller:2008qz}. This conjecture was originally based upon the discovery of a non-trivial Jordan 
cell structure typical for logarithmic CFTs (LCFTs), see e.g.~\cite{Gurarie:1993xq,Flohr:2001zs,Gaberdiel:2001tr}.
Evidence in its favor was later provided by checking 2- and 3-point 
correlators \cite{Skenderis:2009nt,Grumiller:2009mw}. 

One way to decide between these different options is to actually calculate the partition function of TMG 
from first principles, and to compare it to the CFT partition functions of the proposed duals. 
The full partition function $Z$ consists (at least) of two parts, a classical 
contribution $Z_{\rm cl}$ and a 1-loop contribution that we denote by $Z_{\textrm{TMG}}$
\begin{equation}
Z = Z_{\textrm{cl}}\cdot Z_{\textrm{TMG}} \ .
\label{eq:intro1}
\end{equation}
It was argued that the result \eqref{eq:intro1} coincides with the exact result
\cite{Maloney:2007ud,Maloney:2009ck}. Assuming this, the calculation of the full partition function
reduces to a 1-loop calculation for which the tools have already been developed 
\cite{Giombi:2008vd,David:2009xg}\footnote{For earlier papers and further references see 
\cite{Camporesi:1990wm,Camporesi:1994ga,Mann:1996ze,Bytsenko:1994bc}.}. 

In this paper we calculate the 1-loop partition function $Z_{\textrm{TMG}}$ for TMG with 
AdS boundary conditions from first principles. At the critical point we find that 
\begin{equation}
Z_{\textrm{TMG}} = \prod\limits_{n=2}^\infty \frac{1}{|1-q^n|^2}\prod\limits_{m=2}^\infty\prod\limits_{\bar m=0}^\infty \frac{1}{1-q^m\bar q^{\bar m}} \ .
\label{eq:intro2}
\end{equation}
In particular, the expression is not chiral. Furthermore, as we shall explain in detail, it agrees
precisely with what one would expect from a logarithmic conformal field theory of the type proposed in 
\cite{Grumiller:2008qz}. Our calculation gives therefore strong support to the idea that the dual
of TMG at the critical point is indeed logarithmic. 

Finally, we also apply our methods to another 3-dimensional theory, namely new 
massive gravity (NMG) \cite{Bergshoeff:2009hq}. 
\smallskip

This paper is organized as follows: in section \ref{sec:2} we set the stage for the calculation 
of the 1-loop partition function.  The gauge fixing procedure and the calculation of the
ghost determinant is explained in section \ref{sec:ghost}. We then apply these results 
in section~\ref{sec:3} to calculate the 1-loop partition function $Z_{\textrm{TMG}}$. Finally, we discuss 
in section~\ref{sec:4}  our results and compare them with CFT partition functions. The generalization to
NMG is explained in appendix \ref{sec:NMG}, and appendix \ref{sec:B} provides the details of a 
combinatorical argument needed for the interpretation of the dual LCFT.

\section{Preliminaries}\label{sec:2}

The action of cosmological topologically massive gravity for Euclidean signature reads
\begin{equation}
S=\frac 1{\kappa^2} \int d^3x \sqrt{g} \,\left[ R +\frac 2 {\ell^2} +\frac i\mu \,
\varepsilon^{\lambda\mu\nu}\, \Gamma^\rho{}_{\sigma\lambda} \left( \partial_\mu \Gamma^\sigma{}_{\rho\nu}
+\frac 23 \Gamma^\sigma{}_{\kappa\mu}\Gamma^\kappa{}_{\sigma\nu}\right)\right]  \ .
\label{Eact} 
\end{equation} 
The gravitational coupling constant is given by $\kappa^2=16\pi G_N$, where $G_N$ is Newton's constant.
We assume for definiteness that the AdS radius $\ell$ and the Chern--Simons coupling constant $\mu$ 
are both positive. The critical point arises if these constants are tuned as follows.
\begin{equation}
\textrm{Critical\;point:}\qquad \mu\ell=1 \ .
\label{eq:critical}
\end{equation}
The second variation of the action \eqref{Eact} is given by
\begin{equation}
\delta^{(2)}S=-\frac 1{2\kappa^2} \int d^3x \sqrt{g}\, h^{\mu\nu} (D^M L h)_{\mu\nu}\ ,
\label{2nd}
\end{equation}
where $h_{\mu\nu}$ is a small fluctuation around an AdS background
\begin{eqnarray}
&& D^M h_{\mu\nu} =h_{\mu\nu} +\frac i{2\mu} (\tilde Dh)_{\mu\nu} \label{DM}\\
&&(\tilde Dh)_{\mu\nu}= \varepsilon_\mu^{\ \, \rho\beta}\nabla_\rho h_{\nu\beta}
+ \varepsilon_\nu^{\ \, \rho\beta}\nabla_\rho h_{\mu\beta} \label{tilD}\\
&&(Lh)_{\mu\nu}= -\nabla^2 h_{\mu\nu} -\nabla_\mu\nabla_\nu h + \nabla_\nu \nabla^\beta h_{\beta\mu}
+ \nabla_\mu \nabla^\beta h_{\beta\nu} -\frac 2{\ell^2} h_{\mu\nu} \nonumber\\
&&\qquad\qquad - g_{\mu\nu} (\nabla_\rho\nabla_\sigma 
h^{\rho\sigma} - \nabla^2 h)\ .\label{defL}
\end{eqnarray}
The fluctuations $h_{\mu\nu}$ can be decomposed into transverse traceless $h_{\mu\nu}^{TT}$, trace $h$, 
and gauge parts $\nabla_{(\mu}\xi_{\nu)}$
\begin{equation}
h_{\mu\nu}(h^{TT},h,\xi)=h_{\mu\nu}^{TT}+\frac 13 g_{\mu\nu} h + 2\nabla_{(\mu}\xi_{\nu)} \ .
\label{decom}
\end{equation}
By definition $h^{\mu\,TT}_{\;\mu}=\nabla^\mu h_{\mu\nu}^{TT}=0$.  It is easy to show that trace modes
are zero modes of $\tilde D$, gauge modes are zero modes of $L$, and $\tilde D$ maps gauge modes to gauge modes. However, $L$ does not map trace modes to trace modes.

Our aim is to calculate the 1-loop partition function
\begin{equation}
Z_{\textrm{TMG}} = \int {\cal D}h_{\mu\nu}\times\textrm{ghost}\times \exp{\big(-\delta^{(2)}S\big)} \ .
\label{eq:1loop}
\end{equation}
Here `ghost'  refers to all ghost determinants produced by elimination of the gauge degrees of freedom. We shall explain below how these contributions are determined. The path integral is taken over all smooth fluctuations $h_{\mu\nu}$ around the AdS background that are compatible with asymptotic AdS behavior. 
More precisely, we assume that the background geometry is thermal Euclidean AdS$_3$, 
which is the $M_{0,1}$ geometry in the notation of 
\cite{Maloney:2007ud}\footnote{ Note that the full partition function is a sum over partition 
functions $Z_{c,d}(\tau)$ that can be obtained from the one we are calculating, $Z_{0,1}(\tau)$ 
(and which we shall still call $Z_{\rm TMG}$ to keep the notation simple), by modular 
transformations \cite{Maloney:2007ud}.}.
In the 1-loop calculations below this is implicit in the definition of the determinants. Our main 
technical tools to determine the 1-loop partition \eqref{eq:1loop} are zeta-function regularization 
and heat kernel techniques \cite{Vassilevich:2003xt}.

\section{Gauge fixing and ghost determinant}\label{sec:ghost}

The action (\ref{2nd}) is third order in derivatives. Therefore, a suitable gauge-fixing term should
also be third order. Dealing with such gauge-fixing terms is not convenient. Instead,  we use an approach based on an explicit separation of the gauge modes. This method is equivalent in effect to imposing the strong condition
$
\xi_\mu=0 
$
on the fluctuations.  In the context of quantum gravity on de Sitter space such a procedure was used, e.g., in \cite{Vassilevich:1992rk,Mottola:1995sj}, where one can find further references and details.

In terms of the decomposition (\ref{decom})  the quadratic action (\ref{2nd}) takes the form
\begin{equation}
\delta^{(2)}S=-\frac 1{2\kappa^2} \int \!d^3x \sqrt{g}\left[ h^{TT\mu\nu} D^M \big(-\nabla^2 -\frac 2{\ell^2}
\big) h^{TT}_{\mu\nu} +\frac 29 h\big(\nabla^2 -\frac 3{\ell^2} \big)h \right] \ . \label{2gf}
\end{equation}
Due to gauge invariance the action \eqref{2gf} does not contain $\xi$. Thus, the functional integral over gauge degrees of freedom represented by $\xi$ in the measure can be performed trivially. It yields the volume of the gauge group, the diffeomorphism group, which is an infinite constant and has to be eliminated. This works as follows. The path integral measure is divided by the volume of the gauge group to avoid double counting of gauge-equivalent configurations. It is convenient to express this division by the gauge group volume in terms of a (Faddeev--Popov) ghost determinant. The ghost determinant then is given by the Jacobian factor corresponding to the change of variables 
$h_{\mu\nu}\to (h^{TT}_{\mu\nu},h,\xi_\mu)$,
\begin{equation}
\mathcal{D}h_{\mu\nu}=Z_{\textrm{gh}}\, \mathcal{D}h_{\mu\nu}^{TT}\,\mathcal{D}\xi_\mu\,\mathcal{D}h \ .
 \label{defZg}
\end{equation}
The path integral measures for tensor, vector,
and scalar fields are defined by the equations
\begin{eqnarray}
&&1=\int\mathcal{D}h_{\mu\nu}\, \exp (-\langle h,h \rangle )\label{meas2}\\
&&1=\int\mathcal{D}\xi_\mu\, \exp (-\langle \xi,\xi \rangle )\label{meas1}\\
&&1=\int\mathcal{D}\sigma\, \exp (-\langle \sigma,\sigma \rangle ) \ .\label{meas0}
\end{eqnarray}
Here $\langle .,. \rangle$ are ultralocal invariant scalar products
\begin{eqnarray}
&& \langle h,h' \rangle = \int d^3x \sqrt{g} \, h^{\mu\nu}h'_{\mu\nu}\label{hh}\\
&& \langle \xi,\xi' \rangle = \int d^3x \sqrt{g}\, \xi^\mu \xi'_\mu\label{xixi}\\
&& \langle \sigma,\sigma' \rangle = \int d^3x \sqrt{g}\, \sigma\sigma' \ . \label{sisi}
\end{eqnarray}
The vector field $\xi_\mu$ can be decomposed into a transverse $\xi_\mu^T$ and a scalar part $\sigma$
\begin{equation}
\xi_\mu(\xi^T,\sigma) =\xi_\mu^T+\nabla_\mu\sigma \ .
\label{eq:neweq}
\end{equation}
By definition $\nabla^\mu \xi_\mu^T=0$. The Jacobian factor $J_1$ corresponding
to the change of variables $\mathcal{D}\xi_\mu \to \mathcal{D}\xi_\mu^T \mathcal{D}\sigma$
can be calculated from the definition (\ref{meas1})
\begin{eqnarray}
 &&1=\int \mathcal{D}\xi_\mu^T\mathcal{D}\sigma \, J_1\, 
\exp\left(-\int d^3x \sqrt{g} \xi_\nu (\xi^T,\sigma)\xi^\nu(\xi^T,\sigma) \right)\nonumber\\
&&\quad = \int \mathcal{D}\xi_\mu^T\mathcal{D}\sigma \, J_1\, 
\exp\left(-\int d^3x \sqrt{g} (\xi_\nu^T\xi^{T\nu} - \sigma \nabla^2 \sigma ) \right)\nonumber\\
&&\quad =J_1 [\det (-\nabla^2)_0]^{-1/2}\label{J11} \ ,
\end{eqnarray}
where the subscript 0 means that the determinant is calculated for scalar fields. We conclude  that 
\begin{equation}
\mathcal{D}\xi_\mu=J_1\mathcal{D}\xi_\mu^T\mathcal{D}\sigma \qquad
J_1 =[\det (-\nabla^2)_0]^{1/2} \ . \label{J1}
\end{equation}
It is convenient to shift the trace part as $h\to h-2\nabla^2 \sigma$. This change of variables
produces a unit Jacobian factor. The decomposition of the metric then reads as 
\begin{equation}
h_{\mu\nu}(h^{TT},\,h,\,\xi^T,\,\sigma)=h_{\mu\nu}^{TT} +\frac 13 g_{\mu\nu}h +\nabla_\mu\xi_\nu^T
+\nabla_\nu\xi_\mu^T
+2\nabla_\mu\nabla_\nu\sigma -\frac 23 g_{\mu\nu}\nabla^2\sigma \ .\label{decom1}
\end{equation}
The decomposition (\ref{decom1}) is orthogonal with respect to the inner product (\ref{hh}). 
The Jacobian factor induced by the change of the variables $h_{\mu\nu}\to (h^{TT},\,h,\,\xi^T,\,\sigma)$
can be calculated in the same way as above
\begin{equation}
1=\int J_2 \mathcal{D}h^{TT}_{\mu\nu}\mathcal{D}h\mathcal{D}\xi^T_\mu\mathcal{D}\sigma\,
\exp \Big(-\langle h(h^{TT},\,h,\,\xi^T,\,\sigma),\,h(h^{TT},\,h,\,\xi^T,\,\sigma)\rangle\Big) \ ,
\end{equation}
giving
\begin{eqnarray}
&&\mathcal{D}h_{\mu\nu}=J_2 \mathcal{D}h_{\mu\nu}^{TT}\mathcal{D}h\mathcal{D}\xi_\mu^T
\mathcal{D}\sigma \nonumber\\
&&J_2=[\det (-\nabla^2)_0 \det (-\nabla^2 +3/\ell^2)_0 \det (-\nabla^2 +2/\ell^2)_1^T ]^{1/2} \ ,
\label{J33}
\end{eqnarray}
where the subscripts $1$ and $0$ refer to vector and scalar fields, respectively. The ghost factor (\ref{defZg}) in the path integral is then determined by the ratio of the Jacobians (\ref{J33}) and (\ref{J1})
\begin{equation}
Z_{\textrm{gh}}=J_2/J_1=\left[ \det (-\nabla^2 +2/\ell^2)_1^T \det (-\nabla^2 +3/\ell^2)_0\right]^{1/2} \ .
\label{Zgh}
\end{equation}
To derive (\ref{Zgh}) we factorized a determinant of a product of two scalar
operators into a product of two determinants. In the infinite dimensional case
$\det (AB)$ need not coincide with the product $\det(A)\cdot \det(B)$ if one
defines all three determinants independently (e.g., through the zeta function).
However, one can \emph{define} determinants  of higher order operators by reducing them to products of lower order operators. This is the usual rule of
the game in quantum gravity (see \cite{Mottola:1995sj}), which ensures gauge
independence among other nice properties. Besides, even if $\det(AB)$ and
$\det(A)\cdot\det(B)$ are both defined independently through corresponding
zeta functions, the
difference between the determinant of the product and the product of determinants
(the so-called multiplicative anomaly) typically vanishes in odd dimensions
\cite{Elizalde:1997nd}. 

Summarizing our results so far, the partition function reads
\begin{equation}
Z_{\textrm{TMG}}=Z_{\textrm{gh}} \int \mathcal{D}h^{TT}_{\mu\nu}\mathcal{D}h\, \exp(-\delta^{(2)}S) \label{ZZ}
\end{equation}
with the ghost determinant given in \eqref{Zgh} and the quadratic action given in \eqref{2gf}.

Let us now comment on the use of more conventional gauge-fixing procedures.
Consider a gauge-fixing delta function $\delta ((\mathcal{F}h)_\mu)$, where
$\mathcal{F}$ is some operator. In the path integral measure this delta
function has to be accompanied by the compensating determinant $(\det H)$, which
is defined through a linearized gauge transformation of the gauge-fixing condition, 
$H\xi = \mathcal{F}h(h^{TT}=0,h=0,\xi)$. Thus, we have
\begin{eqnarray}
&&Z_{\textrm{TMG}}=\int \mathcal{D} h_{\mu\nu} \, \det (H) \delta ((\mathcal{F}h)_\mu)
\exp(-\delta^{(2)}S)\nonumber\\
&&\qquad =\int Z_{\textrm{gh}} \mathcal{D} h_{\mu\nu}^{TT} \mathcal{D}h\mathcal{D}\xi_\mu
 \, \det (H) \delta ((\mathcal{F}h)_\mu)
\exp(-\delta^{(2)}S) \nonumber\\
&&\qquad =Z_{\textrm{gh}} \int \mathcal{D} h_{\mu\nu}^{TT} \mathcal{D}h\exp(-\delta^{(2)}S)\ ,\nonumber
\end{eqnarray}
where we used (\ref{defZg}) and then integrated over $\xi$. The result, as
expected, coincides with (\ref{ZZ}), which was obtained by more economic methods.


\section{Topologically massive gravity 1-loop partition function}\label{sec:3}

We can now apply the result of the previous section to the case of primary interest, topologically
massive gravity. First we deal with the trace part of the metric. 
With our choice of the overall sign of the action \eqref{2gf} the path integral over 
$h$ in \eqref{ZZ} is convergent without any complex rotations. 
This integral produces a factor of $[\det (-\nabla^2+3/\ell^2)_0]^{-1/2}$.

This leaves us with the contribution from the integration over $h^{TT}$. 
The operator acting 
on the $h^{TT}$ modes in the action (\ref{2gf}) can be factorized as
$D^MD^LD^R$, where $D^{L,R}=1/\ell \pm (i/2)\tilde D$. The operator $\tilde D$
is a symmetric (formally selfadjoint) and even elliptic operator when
restricted to the $TT$ modes. Let $\lambda_j$ denote the eigenvalues of
$\tilde D$. The spectrum of $\{ \lambda_j\}$ is real. The eigenvalues
of $D^MD^LD^R$ are then of the form $\Lambda_j=(1+(i/2\mu)\lambda_j)(1/\ell^2 +\lambda_j^2/4)$,
and hence all $\Lambda_j$ have a positive real part. Therefore, the integral over
each eigenmode converges giving $\Lambda_j^{-1/2}$. Finally, by
collecting the contributions from all $TT$ modes and the trace modes,
we obtain from \eqref{2gf}
\begin{equation}
\int \mathcal{D}h^{TT}_{\mu\nu}\mathcal{D}h\, \exp(-\delta^{(2)}S)=\big[\det (D^M(-\nabla^2-2/\ell^2))_2^{TT}
\det (-\nabla^2 +3/\ell^2)_0 \big]^{-1/2} \ . \label{pin}
\end{equation}
Including the ghost factor (\ref{Zgh}) we then get for the full 1-loop partition function  \eqref{ZZ} of TMG 
\begin{equation}
Z_{\textrm{TMG}}=\sqrt{\frac{\det (-\nabla^2 +2/\ell^2)_1^T}{\det (D^M(-\nabla^2-2/\ell^2))_2^{TT}}} \ .
\label{eq:lalapetz}
\end{equation}
Here the superscript `$T$' in the numerator denotes restriction to transverse vector modes. 

Before we evaluate $Z_{\textrm{TMG}}$ further let us note that for the case of Einstein gravity
the same calculation would have led to 
\begin{equation}
Z_{\textrm{Ein}}=\sqrt{\frac{\det (-\nabla^2 +2/\ell^2)_1^T}{\det(-\nabla^2-2/\ell^2)_2^{TT}}}
\label{ZEin}
\end{equation}
since Einstein gravity is obtained from TMG upon taking the (formal) limit $\mu\to\infty$,
which amounts to neglecting $iD^M$ in (\ref{eq:lalapetz}). The result \eqref{ZEin} 
agrees with what was derived earlier in \cite{Giombi:2008vd,David:2009xg}. This is a nice
consistency check of our method.

Returning to TMG at generic values of $\mu$, we can now factor the determinant in the denominator
of (\ref{eq:lalapetz}),  and write the one-loop partition function as 
\begin{equation}
Z_{\textrm{TMG}}=Z_{\textrm{Ein}}\cdot [\det (iD^M)_2^{TT}]^{-1/2}=Z_{\textrm{Ein}}\cdot Z_M \ . 
\label{ZZEin}
\end{equation}
The fact that the TMG 1-loop partition function contains the Einstein part as a factor makes also sense
physically since TMG has the same fluctuations, ghost- and gauge-fixing terms as in Einstein gravity. 
The remaining factor takes into account massive graviton fluctuations. In order to evaluate this term
let us consider its absolute value
\begin{equation}
|Z_M|=|\det (iD^M)_2^{TT}|^{-1/2}=[\det (D^M\bar D^M)_2^{TT}|^{-1/4}\ , \label{ZMDM}
\end{equation}
where bar means complex conjugation. Now we observe that 
\begin{equation}
\tilde D\tilde Dh_{\mu\nu}=-4\nabla^2h_{\mu\nu}^T 
-2g_{\mu\nu}\nabla^\rho\nabla^\sigma h_{\rho\sigma}^T
+3(\nabla_\mu \nabla^\sigma h_{\nu\sigma}^T
+\nabla_\nu \nabla^\sigma h_{\mu\sigma}^T) 
-\frac {12}{\ell^2} h_{\mu\nu}^T \ , \label{tDtD}
\end{equation}
where $h_{\mu\nu}^T\equiv h_{\mu\nu}-\frac 13 g_{\mu\nu}h$ is the traceless part of the metric 
fluctuations. With the help of the identity \eqref{tDtD} we obtain
\begin{equation}
\det (D^M\bar D^M)_2^{TT}=\det(\mu^{-2} (-\nabla^2 -2/\ell^2 +\delta m^2)_2^{TT}) \ , \label{DMdm}
\end{equation}
where $\delta m^2=\mu^2 -1/\ell^2$.  To evaluate the determinant \eqref{DMdm} we finally use the heat 
kernel methods and the results of \cite{David:2009xg}. According to \cite{Giombi:2008vd,David:2009xg} 
the heat kernel of $(-\nabla^2)_2^{TT}$ reads
\begin{equation}
 K^{(2)}(t)=\sum_{n=1}^\infty \frac {\tau_2\, \cos (2n\tau_1)}{\sqrt{4\pi t}
|\sin (n\tau/2)|^2}
e^{-\frac{n^2\tau_2^2}{4t}}\, e^{-3t} \ , \label{K2}
\end{equation}
where, for convenience, we put $\ell=1$. We also use the same notations and conventions as in  
\cite{David:2009xg}, in particular
\begin{equation}
\tau =\tau_1+i \tau_2\qquad q = \exp{(i\tau)} \ .
\label{eq:notations}
\end{equation}
The quantities $2\pi\tau_1$ and $2\pi\tau_2$ are equivalent to the angular potential $\theta$ and inverse temperature $\beta$, respectively.
The determinant \eqref{DMdm} is then expressed as an integral of the heat kernel.
\begin{equation}
-\ln \det (-\nabla^2 +m^2)_2^{TT}=
\int_0^\infty \frac {dt}t \, K^{(2)}(t)e^{-m^2t} \ , \label{lndetK}
\end{equation}
where the effective mass is, in our case, equal to $m^2=-2+\delta m^2$. The integral over $t$ can 
now be performed straightforwardly 
\begin{equation}
 \frac 1{4\pi^{1/2}} \int\limits_0^\infty \frac {dt}{t^{3/2}} 
\, e^{-\frac{\alpha^2}{4t} -\beta^2 t}=
\frac 1{2\alpha} e^{-\alpha\beta}\label{int}
\end{equation}
yielding
\begin{eqnarray}
&&\ln |Z_M|=\frac 12 \sum_{n=1}^\infty \frac{\cos (2n\tau_1)}{2n |\sin (n\tau/2)|^2}
e^{-n\sqrt{1+\delta m^2}\tau_2}\nonumber\\
&&\qquad\qquad =\sum_{n=1}^\infty \frac 1{2n} |q|^{n(|\mu |-1)}\, \frac{q^{2n}
+\bar q^{2n}}{(1-q^n)(1-\bar q^n)} \ .
\label{absZM}
\end{eqnarray}
The dependence of the determinant (\ref{DMdm}) on the overall scale $\mu^{-2}$ is given by the global
scale anomaly. In the $\zeta$-regularization this anomaly is the $t^0$ term in the short-$t$ asymptotics
of the heat kernel, and this term vanishes in our case. Therefore, (\ref{absZM}) is indeed the final
answer for the absolute value of the contribution to the partition function $Z_M$.

Knowing the absolute value of the partition function $Z_M$ we can now make an educated guess 
for its phase.  We confine ourselves to the critical point \eqref{eq:critical}.
Rewriting $\ln |Z_M|=\frac 12\ln{Z_M}+\frac 12\ln{\bar Z_M}$ it is suggestive from the second line of 
\eqref{absZM} that the correct result for the partition function $Z_M$ at the critical point is given by
\begin{equation}
\ln Z_M = \sum_{n=1}^\infty \frac 1{n}\,\frac{q^{2n}}{(1-q^n)(1-\bar q^n)}\qquad\textrm{for}\; \mu\ell=1 \ .
 \label{eq:ZM}
\end{equation}
Even at the critical point the TMG 1-loop partition function \eqref{eq:ZM} depends on both
$q$ and $\bar q$. It is therefore not chiral as opposed to the classical `chiral gravity' partition function 
\cite{Maloney:2009ck}
\begin{equation}
Z_{\rm cl} = \bar q^{-\ell/8G_N}\qquad\textrm{for}\; \mu\ell=1 \ .
 \label{eq:Zcl}
\end{equation}

The fact that the massive graviton determinant \eqref{absZM} is non-trivial may seem 
surprising, since in Euclidean signature there are no classical solutions leading to real metrics 
that contain massive graviton excitations \cite{Maloney:2009ck}. In other words, the zero spectrum 
of the operator $D^M D^L D^R$  consists only of those (real) modes that are annihilated by 
$D^L D^R$. Thus, the Wick rotation from Minkowski to Euclidean signature eliminates many classical 
solutions if we insist on real metrics. However, in the quantum case we have to find an eigenspectrum 
of the same operator in the space of real square-integrable fluctuations. The latter condition yields
complex eigenvalues, but the right number of them, i.e., the same number as in the Minkowski case.  
In this sense the Wick rotation may be ill-defined classically, but well-defined for 1-loop calculations. 

At the critical point \eqref{eq:critical} there is also another subtlety that should be mentioned. 
It concerns the split of the determinant $\det{(D^LD^LD^R)}=\det (D^L)\, \det (D^L D^R)$, which,
as was mentioned before, may not hold for operators on infinite dimensional spaces.
In particular, zero modes have to be excluded from the determinants and treated
separately, and this may spoil the factorization. This is precisely what happens
in the Minkowski signature TMG at the critical point,
but does not appear for the Euclidean signature, as we have discussed above.
Note, that one does not expect in general a mode-by-mode correspondence
between the Minkowski and Euclidean spectra, though there is a correspondence
at the level of partition functions\footnote{A good example of such kind is field
theories at finite temperature.}. Another potential source of troubles are the
UV divergences in the determinants. However, in odd dimensions the zeta-regularized
determinants are finite, and, therefore, the multiplicative anomaly vanishes
\cite{Elizalde:1997nd}. 

In any case, we shall now assume that \eqref{eq:ZM} is the correct 1-loop partition function
at the chiral point, and study the implications of it. 
In particular, we want to show that \eqref{eq:ZM}  has a very natural interpretation in terms 
of a dual LCFT as originally proposed in \cite{Grumiller:2008qz}.

\section{Discussion and comparison with CFT partition functions}\label{sec:4}

In order to explain the relation between the 1-loop partition function and the partition function
of the dual conformal field theory let us first recall how this worked for the case of Einstein 
gravity. For Einstein gravity we need to evaluate the determinants that appear in  \eqref{ZEin}. 
In addition to the formula for the spin two heat kernel (\ref{K2}) we also need the expression for
the vector hear kernel \cite{Giombi:2008vd,David:2009xg}
\begin{equation}
 K^{(1)}(t)=\sum_{n=1}^\infty 
\frac {\tau_2\, \cos (n\tau_1)}{\sqrt{4\pi t} |\sin (n\tau/2)|^2}
e^{-\frac{n^2\tau_2^2}{4t}}\, e^{-2t}  \ ,
\label{eq:K1}
\end{equation}
where we have used the same conventions as above. Altogether this then leads to 
\begin{equation}
\ln{Z_{\textrm{Ein}}} = \sum_{n=1}^\infty  \frac{\cos{(2n\tau_1)}e^{-n\tau_2}-\cos{(n\tau_1)}
e^{-2n\tau_2}}{2n|\sin{(n\tau/2)}|^2} = \sum_{n=1}^\infty\frac1n\,\Big(\frac{q^{2n}}{1-q^n}
+\frac{\bar q^{2n}}{1-\bar q^n}\Big) \ .
\label{eq:Einein}
\end{equation}
Exponentiating the result \eqref{eq:Einein} and using the power series expansion of
$\ln (1 - q^n)$ we then obtain
\begin{equation}
Z_{\textrm{Ein}} = \prod_{n=2}^\infty \frac{1}{|1-q^n|^2} \ .
\label{eq:ZEinein}
\end{equation}
{}From the dual conformal field theory point of view, this is just the partition function of 
the Virasoro vacuum representation. Note that (\ref{eq:ZEinein}) factorizes holomorphically,
although this is not manifest from the first equation in (\ref{eq:Einein}). Indeed, it is only 
the combined contribution from the tensor and the vector part (the difference of the two terms in 
the first equation in (\ref{eq:Einein})) that factorizes.

At the critical point $\mu\ell=1$, we can similarly determine the contribution to the partition function
coming from $Z_M$. Using 
\begin{eqnarray}
\sum_{n=1}^{\infty} \frac{1}{n} \, \frac{q^{2n}}{(1-q^n)(1-\bar{q}^n)}  & = &
\sum_{m=0}^{\infty} \sum_{\bar{m}=0}^{\infty}  \, 
\sum_{n=1}^{\infty} \frac{1}{n} \, q^{2n+mn} \bar{q}^{n\bar{m}} \nonumber \\
& = & - \sum_{m=2}^{\infty} \sum_{\bar{m}=0}^{\infty}  \log (1 - q^{m} \bar{q}^{\bar{m}}) \ ,
\end{eqnarray}
where we have used the geometric series expansion in the first line, we obtain
\begin{equation}
 Z_{\rm TMG} = \prod\limits_{n=2}^\infty \frac{1}{|1-q^n|^2}
 \prod\limits_{m=2}^\infty\prod\limits_{\bar m=0}^\infty \frac{1}{1-q^m\bar q^{\bar m}} \qquad\textrm{for}\;
  \mu\ell=1 \ . 
 \label{eq:ZTMG}
\end{equation}
Note that this partition function does {\em not} factorize holomorphically. Technically this comes from the fact
that now there is only a tensor contribution to the determinant $Z_M$, but no vector contribution as for Einstein gravity. The absence 
of an additional vector part coming from the gauge symmetries is not surprising, but merely a consequence 
of the fact that we have only one set of diffeomorphisms that can act on our fluctuations. 
\medskip

The complete 1-loop partition function (\ref{eq:ZTMG}) 
 should now be compared to the partition function of the logarithmic conformal
field theory proposed in \cite{Grumiller:2008qz}. The left-moving central charge
of this conformal field theory vanishes, $c_L=0$, and the right-moving central charge is non-vanishing, $c_R=3\ell/G_N$. The left-moving stress energy tensor 
$T(z)$ has a logarithmic partner $t(z)$, satisfying
\begin{equation}
L_0\, t = 2\, t + T \ , \qquad  L_0 T = 2\, T \ , \qquad L_1 t = L_1  T = 0 \ .
\end{equation}
Furthermore, the two-point function $\langle T(z)\, t(w) \rangle = \tfrac{b_L}{(z-w)^4}$ is non-zero, which 
implies that 
\begin{equation}
L_2 \, t = b_L\, \Omega \ , \qquad b_L=-c_R=-\frac{3\ell}{G_N} \ ,
\end{equation}
where $\Omega$ is the ground-state vacuum of the LCFT. Here we have
used that $T=L_{-2}\Omega$ which also implies that $L_2 T = 0$. Moreover, $T$ and $t$ are annihilated by
all positive $\bar{L}_n$ modes, as well as by all modes $L_n$ with $n\geq 3$. Finally, the 
consistency of the LCFT,
in particular locality, implies that $L_0-\bar{L}_0$ is diagonalizable, and thus
\begin{equation}
\bar{L}_0 \, t = T \ .
\end{equation}
The structure of the low-lying states can therefore be summarized by the diagram 
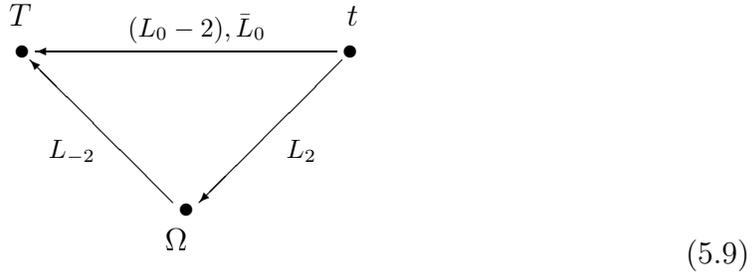
\begin{equation}
 \begin{picture}(140,120)(-10,-20)
      \put(0,60){\vbox to 0pt
        {\vss\hbox to 0pt{\hss$\bullet$\hss}\vss}}
      \put(124,60){\vbox to 0pt
        {\vss\hbox to 0pt{\hss$\bullet$\hss}\vss}}
      \put(62,0){\vbox to 0pt
        {\vss\hbox to 0pt{\hss$\bullet$\hss}\vss}}
      \put(119,60){\vector(-1,0){114}}
      \put(57,3){\vector(-1,1){54}}
      \put(121,57){\vector(-1,-1){54}}
      \put(-5,70){$T$}
      \put(123,70){$t$}
      \put(40,66){{\footnotesize $(L_0-2),\bar{L}_0$}}
      \put(54,-15){$\Omega$}
       \put(10,20){{\footnotesize $L_{-2}$}}
        \put(100,20){{\footnotesize $L_{2}$}}
    \end{picture}
 \end{equation}
In addition there is the right-moving stress energy tensor $\bar{T}=\bar{L}_{-2}\Omega$
that satisfies the usual properties of the holomorphic flux component of a CFT stress energy tensor.

In order to determine the contribution of the above states to the partition function
we first consider the subset of those states that are descendants of $\Omega$. As is
clear from the above diagram, they are unaffected by the presence of $t$, and are hence
counted by the usual Virasoro partition function
\begin{equation}
 Z_{\Omega} =  \prod\limits_{n=2}^\infty \frac{1}{|1-q^n|^2} \ .
 \label{eq:Zvirasorodesc}
\end{equation}
The remaining states are therefore descendants of $t$ that are not already descendants of
$\Omega$. It follows from the above relations that the positive modes $L_n$ and $\bar{L}_n$
with $n> 0$ either annihilate $t$, or map it to $\Omega$. Furthermore, $(L_0-2)$ and $\bar{L}_0$
map $t$ to a descendant of $\Omega$, namely $T=L_{-2}\Omega$. Thus we only need to consider
the descendants of $t$ by negative modes. Since $t$ is a logarithmic mode it is neither annihilated by 
$L_{-1}$ nor by $\bar{L}_{-1}$. Thus the additional contribution is simply 
\begin{equation}
 Z_{\rm t} =  q^2\,\prod\limits_{n=1}^\infty \frac{1}{|1-q^n|^2} \ ,
 \label{eq:Zlogdesc}
\end{equation}
where the overall factor of $q^2$ comes from the fact that $t$ has eigenvalue $(2,0)$ under
the diagonal part of $(L_0,\bar{L}_0)$. 

The total partition function of the Virasoro descendants of the above states is then 
\begin{equation}
Z^0_{\rm LCFT} = Z_{\Omega} +  Z_{\rm t} = 
 \prod\limits_{n=2}^\infty \frac{1}{|1-q^n|^2}\,\Bigl(1 + \frac{q^2}{|1-q|^2} \Bigr) \ .
 \label{eq:ZLCFT}
\end{equation}
This is now to be compared with the TMG partition function \eqref{eq:ZTMG}. To see how the
two results fit together we first observe that the first factor that equals $Z_{\Omega}$ agrees in both
cases. It describes the contribution of the usual boundary gravitons corresponding to $T$ and
$\bar{T}$. Note that the term proportional to $q^{nl}$ in 
\begin{equation}\label{inter}
\frac{1}{1-q^n} = \sum_{l=0}^{\infty} q^{nl} 
\end{equation}
denotes the contribution of the multi-graviton corresponding to $L_{-n}^l$, and similarly for 
$\tfrac{1}{1-\bar{q}^n}$. 

The remaining factor in \eqref{eq:ZTMG} should therefore describe
the multi-particle excitations corresponding to $t$. On the LCFT side we do not reproduce
this full factor since the LCFT partition function \eqref{eq:ZLCFT} only takes into account Virasoro
descendants of $t$, but does not include any multi-particle $t$-excitations. Indeed, $t$ is not really
part of the chiral algebra (or vertex operator algebra) since it is a non-chiral logarithmic field,
and as a consequence we do not know how to introduce modes for it or calculate the
commutation relations, etc. Thus the LCFT
partition function can only be compared to the {\em single-particle} $t$-excitations of \eqref{eq:ZTMG}. 
In the spirit of (\ref{inter}) these are the terms where we only keep the linear term of each
denominator,\footnote{Alternatively, these terms are just the $n=1$ contribution of 
(\ref{eq:ZM}).}  i.e.\
\begin{equation}\label{split}
\prod\limits_{m=2}^\infty\prod\limits_{\bar m=0}^\infty \frac{1}{1-q^m\bar q^{\bar m}} 
= 1 + \sum_{m=2}^{\infty} \sum_{\bar{m}=0}^{\infty} q^m \bar{q}^{\bar{m}} + \hbox{multiparticle.} 
\end{equation}
The single particle term now agrees perfectly with \eqref{eq:ZLCFT} since
\begin{equation}
\Bigl(1 + \frac{q^2}{|1-q|^2} \Bigr)  = 1 + q^2 \sum_{m=0}^{\infty} \sum_{\bar{m}=0}^{\infty} q^m \bar{q}^m \ .
\end{equation}
Obviously, the multiparticle terms in (\ref{split}) also have a good LCFT interpretation: they describe
additional (non-chiral) representations one has to add to the theory in order to make it consistent. (In particular,
these will be the states that are produced in OPEs of $t$ with itself, etc.) This is indeed consistent
since we can write 
\begin{equation}
Z_{\rm TMG} = \prod\limits_{n=2}^\infty \frac{1}{|1-q^n|^2}\, 
\prod\limits_{m=2}^\infty\prod\limits_{\bar m=0}^\infty \frac{1}{1-q^m\bar q^{\bar m}} 
 =  Z^0_{\rm LCFT}
+ \sum_{h,\bar{h}} N_{h,\bar{h}}\, q^h \bar{q}^{\bar{h}} \, \prod_{n=1}^{\infty} \frac{1}{|1-q^n|^2} \ ,
\label{eq:angelinajolie}
\end{equation}
where the last term describes the character of the $(h,\bar{h})$ representation of the Virasoro algebra,
and $N_{h,\bar{h}}$ is the multiplicity with which this representation occurs. 
We have checked explicitly that the first few coefficients $N_{h,\bar h}$ are indeed non-negative 
integers, see Table 1. This can also be done analytically; a simple combinatorial argument is sketched in 
appendix~\ref{sec:B}. 
{\footnotesize
\TABULAR{|r|cccccccccccccc|}{
\hline
$\bar h=$ & 0 & 1 & 2 & 3 & 4 & 5 & 6 & 7 & 8 & 9 & 10 & 11 & 12 & 13 \\ \hline
$h \leq 3$: & 0 & 0 & 0 & 0 & 0 & 0  & 0 & 0 & 0 & 0 & 0 & 0 & 0 & 0  \\
$h=4$: & 1 & 0 & 1 & 0 & 1 & 0 & 1 & 0 & 1 & 0 & 1 & 0 & 1 & 0 \\
$h=5$: & 0 & 1 & 0 & 1 & 0 & 1 & 0 & 1 & 0 & 1 & 0 & 1 & 0 & 1 \\
$h=6$: & 2 & 0 & 2 & 1 & 2 & 1 & 3 & 1 & 3 & 2 & 3 & 2 & 4 & 2 \\
$h=7$: & 0 & 2 & 1 & 2 & 2 & 3 & 2 & 4 & 3 & 4 & 4 & 5 & 4 & 6 \\
$h=8$: & 3 & 1 & 4 & 3 & 6 & 4 & 8 & 6 & 10 & 8 & 12 & 10 & 15 & 12 \\
$h=9$: & 1 & 3 & 3 & 6 & 5 & 9 & 9 & 12 & 12 & 17 & 16 & 21 & 21 & 26\\
$h=10$: & 4 & 3 & 8 & 7 & 14 & 13 & 20 & 20 & 29 & 28 & 39 & 38 & 50 & 50\\
$h=11$: & 2 & 6 & 7 & 13 & 15 & 22 & 26 & 35 & 39 & 51 & 56 & 70 & 77 & 93\\
$h=12$: & 7 & 5 & 15 & 17 & 29 & 32 & 50 & 53 & 76 & 83 & 109 & 119 & 153 & 163 \\
$h=13$: & 3 & 11 & 15 & 26 & 35 & 52 & 64 & 89 & 106 & 138 & 163 & 203 & 234 & 287 \\
$h=14$: & 10 & 11 & 27 & 35 & 60 & 73 & 111 & 132 & 183 & 216 & 283 & 328 & 417 & 476 \\
$h=15$: & 7 & 17 & 29 & 52 & 73 & 111 & 148 & 203 & 259 & 341 & 418 & 529 & 638 & 783 \\
$h=16$: & 14 & 20 & 48 & 67 & 118 & 154 & 234 & 298 & 416 & 513 & 681 & 824 & 1052 & 1252 \\
$h=17$: & 11 & 30 & 53 & 97 & 146 & 225 & 314 & 442 & 582 & 781 & 992 & 1275 & 1581 & 1976 \\
\hline
}{Coefficients $N_{h,\bar h}$ for $h < 18$ and $\bar h < 14$}
}

The heat kernel 1-loop calculation in TMG thus leads to a result that is perfectly consistent with
the proposal of \cite{Grumiller:2008qz} that the dual conformal field theory is logarithmic. This provides
strong support in favor of this claim.

\appendix

\section{Generalization to new massive gravity}\label{sec:NMG}

The analysis for TMG can also be easily generalized to the case of NMG, as we shall now explain. 
At the critical point --- the tuning of parameters in the action where both central charges vanish --- the 
quadratic action for NMG reads \cite{Bergshoeff:2009hq}
\begin{equation}
\delta^{(2)}S_{\textrm{NMG}}=\frac 1{m^2} \int d^3x \sqrt{g}
\left( h^{\mu\nu} \mathcal{G}_{\mu\nu}(k) +\frac 14
(k^{\mu\nu}k_{\mu\nu}-k^2) \right)\ , \label{eq:2NMG}
\end{equation}
where $\mathcal{G}_{\mu\nu}$ is the linearized Einstein tensor including the
cosmological constant, $\mathcal{G}_{\mu\nu}(k)=(Lk)_{\mu\nu}$, see (\ref{defL}).
The choice of the overall sign in Euclidean space will become clear below.
Let us expand the metric fluctuations $h_{\mu\nu}$ and the auxiliary field
fluctuations $k_{\mu\nu}$ in the $TT$, vector, and trace parts
\begin{equation}
k_{\mu\nu}=k_{\mu\nu}^{TT}+\frac 13 g_{\mu\nu} \bar k +2\nabla_{(\mu}v_{\nu)} \ ,
\label{eq:exk}
\end{equation}
cf.~(\ref{decom}) for the corresponding expansion of $h_{\mu\nu}$. Note, that there 
is no gauge symmetry corresponding to $v_\mu$, 
and that $k=k_\mu^\mu =\bar k+2\nabla_\mu v^\mu$. 
The path integral measures can then be expanded as in (\ref{defZg}) with $Z_{\textrm{gh}}$ given by 
(\ref{Zgh}), i.e.
\begin{equation}
\mathcal{D}h_{\mu\nu}\mathcal{D}k_{\mu\nu}=Z_{\textrm{gh}}^2
\mathcal{D}h_{\mu\nu}^{TT} \mathcal{D}h\mathcal{D}\xi_\mu
\mathcal{D}k_{\mu\nu}^{TT} \mathcal{D}\bar k\mathcal{D}v_\mu \ . \label{eq:dechk}
\end{equation}
The first term in the action (\ref{eq:2NMG}) reads similar to (\ref{2gf})
\begin{equation}
\int d^3x \sqrt{g}  h^{\mu\nu} \mathcal{G}_{\mu\nu}(k)\nonumber\\
=
\int d^3x \sqrt{g}\left[ h^{TT\mu\nu}  \big(-\nabla^2 -\frac 2{\ell^2}
\big) k^{TT}_{\mu\nu} +\frac 29 h\big(\nabla^2 -\frac 3{\ell^2} \big)\bar k
\right] \ . \label{eq:2hkact}
\end{equation}
The integration over $h^{TT}$ and $h$ produces delta functions for
$k^{TT}$ and $\bar k$, together with corresponding determinants. The integration
over the gauge modes $\xi_\mu$ is done trivially. One then finds 
\begin{eqnarray}
Z_{\textrm{NMG}}(\textrm{crit}) & = & \int \mathcal{D}h_{\mu\nu}\mathcal{D}k_{\mu\nu} 
\exp ( -\delta^{(2)}S_{\textrm{NMG}}) \label{eq:Z1nmg} \\
& = & Z_{\textrm{gh}}^2 \left[\det\big(-\nabla^2 -\frac 2{\ell^2}
\big)_2^{TT} \det \big(-\nabla^2 +\frac 3{\ell^2} \big)_0 \right]^{-1}
\int\mathcal{D}v_\mu e^{-S(v)} \ , \nonumber  
\end{eqnarray} 
where $S(v)$ is the second term in (\ref{eq:2NMG}) with the substitution
$k_{\mu\nu}\to 2\nabla_{(\mu}v_{\nu)}$. It is convenient to decompose
$v_\mu$ in transversal and longitudinal parts as 
\begin{equation}
v_\mu=v_\mu^T+\nabla_\mu u \qquad \nabla^\mu v_\mu^T=0\ .\label{eq:decv}
\end{equation}
The measure transforms as in (\ref{J1}),
\begin{equation}
\mathcal{D}v_\mu=J_1\mathcal{D}v_\mu^T\mathcal{D}u \qquad
J_1=[\det(-\nabla^2)_0]^{1/2}\ ,
\label{eq:Juv}
\end{equation}
while the action reads
\begin{equation}
S(v)=\frac 1{2m^2} \int d^3x \sqrt{g} \big( v^{T\mu} (-\nabla^2 +2/\ell^2)v_\mu^T
-(4/\ell^2) u \nabla^2 u \big) \ .\label{eq:Sv}
\end{equation}
With these preparations we can now perform the remaining  integrations. Note that 
with our choice of the overall sign in the action (\ref{eq:2NMG}) 
both integrals, over $v^T$ and $u$, are of decaying exponents. 
The integration over $u$ produces a scalar determinant which cancels the Jacobian 
factor $J_1$ in  (\ref{eq:Juv}). 
The integral over $v^T$ yields
\begin{equation}
Z_1=[\det (-\nabla^2 +2/\ell^2)_1^T]^{-1/2}  \ . \label{eq:Z1}
\end{equation} 
Collecting together everything we then obtain
\begin{equation}
Z_{\textrm{NMG}}(\textrm{crit})=Z_{\textrm{Ein}}^2\cdot Z_1= Z_{\textrm{Ein}} 
\cdot [\det (-\nabla^2 -2/\ell^2)_2^{TT}]^{-1/2} \ .  \label{eq:ZNMG}
\end{equation}
At the critical point the NMG 1-loop partition function is thus reduced to the product of the 
Einstein gravity 1-loop partition function times the tensor determinant we have calculated 
above, see \eqref{DMdm}-\eqref{absZM} for $\delta m=0$ (again we set $\ell=1$) 
\begin{equation}
\ln{[\det (-\nabla^2 -2)_2^{TT}]^{-1/2}} = \sum_{n=1}^\infty \frac 1{n} \, 
\frac{q^{2n}+\bar q^{2n}}{(1-q^n)(1-\bar q^n)}  \ .
\label{eq:detNMG}
\end{equation}
The full NMG partition function at the critical point then reads
\begin{equation}
Z_{\textrm{NMG}}(\textrm{crit})= \prod\limits_{n=2}^\infty \frac{1}{|1-q^n|^2}
 \prod\limits_{m=2}^\infty\prod\limits_{\bar m=0}^\infty \frac{1}{1-q^m\bar q^{\bar m}} \prod\limits_{l=0}^\infty\prod\limits_{\bar l=2}^\infty \frac{1}{1-q^l\bar q^{\bar l}} \ .
\label{eq:ZNMGfinal}
\end{equation}
The result \eqref{eq:ZNMGfinal} can now be compared with the partition function of the 
LCFT dual,\footnote{The conjecture that there is a LCFT dual for NMG at the chiral point also 
is supported by the calculation of 2-point correlators on the gravity side 
\cite{Grumiller:2009sn,Alishahiha:2010bw}.} 
\begin{equation}
 Z_{\textrm{LCFT}}^{\textrm{NMG}} = \prod_{n=2}^{\infty} \frac{1}{|1-q^n|^2} \,
\left( 1 + \frac{q^2 + \bar{q}^2}{|1-q|^2} \right)
 \label{eq:lasteq}
\end{equation}
in complete analogy to the discussion
in section \ref{sec:4}. Again all multiplicity coefficients $N_{h,\bar h}$ in the expression 
that is analogous to \eqref{eq:angelinajolie} turn out to be positive. The corresponding 
combinatorial counting argument is essentially the same as the one presented in appendix \ref{sec:B} below.
This provides a fairly
non-trivial check on the validity of the LCFT conjecture for NMG at the critical point.

\section{A combinatorial counting argument}\label{sec:B}

In this appendix we show that the coefficients $N_{h,\bar{h}}$ defined in (\ref{eq:angelinajolie}) 
are indeed non-negative integers. We begin by considering the function
\begin{equation}
D=\prod_{m=2}^{\infty} \prod_{\bar{m}=0}^{\infty} \frac{1}{(1-q^m\bar{q}^{\bar{m}})} 
= 1+ \sum_{h,\bar{h}} B(h,\bar{h}) q^h \bar{q}^{\bar{h}} \ ,
\end{equation}
whose Fourier coefficients $B(h,\bar{h})$ are manifestly non-negative. Indeed, they count 
pairs of partitions, where $h$ is partitioned into integers greater or equal to $2$, while 
$\bar{h}$ is partioned into positive integers, with the constraint that the number of terms
in the partition of $h$ is bigger or equal than that in the partition of $\bar{h}$. Next we consider 
\begin{equation}
\tilde D= D\, (1-q)(1-\bar q) = 1+ \sum_{h,\bar{h}} \tilde B(h,\bar{h})\,  q^h \bar{q}^{\bar{h}} \ ,
\end{equation}
whose Fourier coefficients satisfy by construction
\begin{equation}
\tilde{B}(h,\bar h)=B(h,\bar{h}) - B(h-1,\bar{h}) - B(h,\bar{h}-1) + B(h-1,\bar{h}-1) \ .
\end{equation}
At least for  $h\geq 3$, $\bar{h}\geq 2$, it then follows that also the coefficients 
$\tilde{B}(h,\bar h)$ are non-ngeative, since for every pair of partitions counted by 
$B(h-1,\bar{h})$ and $B(h,\bar{h}-1)$, there is a partition counted by $B(h,\bar{h})$, and
the partitions that arise simultaneously from both $B(h-1,\bar{h})$ and $B(h,\bar{h}-1)$ are 
counted by  $B(h-1,\bar{h}-1)$. The function that appears in (\ref{eq:angelinajolie}) differs from
$\tilde{D}$ by some low order terms, so the above argument proves that the coefficients 
$N_{h,\bar{h}}$ are non-negative for $h\geq 3$ and $\bar{h}\geq 2$. Together with the 
explicit formulas in table~1 this then proves that all $N_{h,\bar{h}}$ are indeed non-negative.

\acknowledgments

We thank Simone Giombi, Rajesh Gopakumar, Niklas Johansson, Andy Strominger and Roberto Volpato for discussions.

DG and DV were supported by the START project Y435-N16 of the Austrian Science Foundation (FWF). DV was supported in part by CNPq and FAPESP. The research of MRG is supported in parts by the Swiss National Science Foundation.
MRG and DV also acknowledge financial support from the Erwin-Schr\"odinger
Institute (ESI) during the workshop ``Gravity in three dimensions''.  MRG thanks Harvard University for
hospitality during the final stages of this work.



\providecommand{\href}[2]{#2}\begingroup\raggedright\endgroup

\end{document}